\documentclass[preprint,preprintnumbers,amsmath,amssymb]{revtex4}
\usepackage{graphicx}% Include figure files
\usepackage{dcolumn}% Align table columns on decimal point
\usepackage{bm}% bold math

\begin{document}

%\preprint{PREPRINT L-2}

\title{$\gamma p$ and $\gamma \gamma$ scattering from $\bar{p}p$, $pp$ forward
amplitudes in a QCD eikonal model with a dynamical gluon mass} % Force line breaks with \\

\author{E.G.S. Luna$^{1,2}$ and A.A. Natale$^{1}$}
%\altaffiliation[Also at ]{Physics Department, XYZ University.}%Lines break automatically
%or can be forced with \\
\affiliation{
$^{1}$Instituto de F\'{\i}sica Te\'orica, \\
UNESP, S\~ao Paulo State University, 01405-900, S\~ao Paulo, SP, Brazil \\
$^{2}$Instituto de F\'{\i}sica Gleb Wataghin, \\
Universidade Estadual de Campinas,
13083-970, Campinas, SP, Brazil}


\begin{abstract}
We examine the $\gamma p$ photoproduction and the hadronic $\gamma \gamma$ total cross sections
by means of a QCD eikonal model with a dynamical infrared mass scale. In this model,
where the dynamical gluon mass is the natural regulator for the tree level gluon-gluon scattering,
the $\gamma p$ and $\gamma \gamma$ total cross sections are derived
from the $pp$ and $\bar{p}p$ forward scattering amplitudes assuming vector meson dominance
and the additive quark model. We show that the validity of the cross section factorization relation
$\sigma_{pp}/\sigma_{\gamma p}=\sigma_{\gamma p}/\sigma_{\gamma \gamma}$ is
fulfilled depending on the Monte Carlo model
used to unfold the hadronic $\gamma \gamma$ cross section data,
and we discuss in detail the case of $\sigma (\gamma \gamma \to hadrons)$ data 
with $W_{\gamma \gamma} > 10$ GeV unfolded by the Monte Carlo generators PYTHIA and PHOJET.
The data seems to favor a mild dependence with the energy of the probability ($P_{had}$)
that the photon interacts as a hadron. 
\end{abstract}


\maketitle

\section{Introduction}

The hadronic nature of the photon is currently a subject of intense theoretical and
experimental interest. The central question for hadronic interactions of real
photons is whether they behave the same as hadrons or whether the additional hard contributions
to the total cross sections of photon-induced interactions lead to a faster rise of the total
$\gamma \gamma$ and $\gamma p$ cross sections as a function of the energy.

The increase of hadron-hadron total cross sections was theoretically predicted many years
ago \cite{cheng} and this prediction has been accurately verified by experiment \cite{eidelman}. The
measurements of the total photoproduction cross sections $\sigma^{\gamma p}$ at HERA \cite{zeus,h1} and
the measurements of the total hadronic cross sections $\sigma^{\gamma \gamma}$ at
LEP \cite{acciarri,abbiendi} have also established the increase of photon-hadron and photon-photon
total cross sections. Early modeling of these cross sections within Regge theory shows a
energy dependence similar to the ones of nucleon-nucleon \cite{land1}. This universal
behavior, appropriately
scaled in order to take into account the differences between hadrons and the photon, 
can be understood as follows: at high center-of-mass energies the total
photoproduction $\sigma^{\gamma p}$ and the total hadronic cross section $\sigma^{\gamma \gamma}$
for the production of hadrons in the interaction of
one and two real photons, respectively, are expected to be dominated by
interactions where the photon has fluctuated into a
hadronic state. Therefore measuring the energy dependence of photon-induced processes should improve our
understanding of the hadronic nature of the photon as well as the universal high energy
behavior of total hadronic cross sections.

However the comparison of the experimental data and the theoretical prediction, as discussed some time ago
\cite{rembold}, may also present some subtleties depending on the Monte Carlo model used to analyze
the data.
Despite the similar energy dependence of hadron-hadron, photon-hadron and photon-photon total cross
sections, there are significant differences between photon induced cross section measurements and those
of the purely hadronic processes. The latter comes from a counting rate $\Delta N(t)$, the number
of counts/sec/$\Delta t$ in a narrow interval around the four-momentum transfer squared $t$, extrapolated
to the value $\Delta N(t=0)$ and normalized by an appropriate factor (for colliding beams this
normalization factor is the luminosity $L$). The $\gamma \gamma$ cross sections are extracted from
a measurement of hadron production in $e^{+}e^{-}$ processes and are strongly dependent upon the acceptance
corrections to be employed. These corrections are in turn sensitive to the Monte Carlo models used in the
simulation of the different components of an event, and this general procedure
produces uncertainties in the determination of $\sigma^{\gamma \gamma}$ \cite{rembold}. This clearly implies
that any phenomenological analysis has to take properly into account the discrepancies among
$\sigma^{\gamma \gamma}$ data obtained from different Monte Carlo generators.

A global description of photon-photon, photon-hadron and hadron-hadron total cross sections is possible by
means of QCD-inspired eikonal models \cite{gregores,godbole,corsetti,durand1}. In these models the increase
of the total
cross sections is associated with semihard scatterings of partons in the hadrons. The energy dependence of
the cross sections is driven mainly by gluon-gluon scattering processes, since $g(x,Q^{2})\gg q(x,Q^{2})$
at small-$x$ values. Nevertheless, the gluon-gluon subprocess cross section is potentially divergent at small
transferred momenta, and the usual procedure to regulate this behavior is the introduction of a purely
{\it ad hoc} parameter separating the perturbative from the non-perturbative QCD
region, like an infrared mass scale \cite{margolis1,block7,gregores}, or a cut-off at low transverse
momentum $p_{T}$ \cite{godbole,corsetti,durand1,durand}. These arbitrary parameters are adjusted in order to
obtain the best fits to the experimental data.

Recently we introduced a QCD-inspired eikonal model where the {\it ad hoc} infrared mass scale was substituted by
a dynamical gluon mass one \cite{luna01}. One of the central advantages of the model is that it gives a precise
physical meaning for the quoted infrared scale. Furthermore, since the behavior of the running coupling constant
is constrained by the value of dynamical gluon mass \cite{aguilar}, the model also has a smaller number of
parameters than similar models.

In this work we perform a detailed study of the $\gamma p$ and $\gamma \gamma$ scattering in the framework of the
QCD eikonal model with a gluon dynamical mass of Ref. \cite{luna01}. We address the question of derivating
$\gamma p$ and $\gamma \gamma$ total cross sections from the $pp$ and $\bar{p}p$ forward amplitudes,
exploring the process underlying the extraction of $\gamma \gamma$ from the $e^{+}e^{-}$ data.
The paper is
organized as follows: in the next section we briefly review the QCD-inspired eikonal model with a dynamical
gluon mass. Sec. III contains the details of $\gamma \gamma$ and $\gamma p$ cross-sections calculations. The results
are presented in the Sec. IV, where we provide fits for the cross sections, discuss the
factorization theorem, $\sigma_{pp}/\sigma_{\gamma p}=\sigma_{\gamma p}/\sigma_{\gamma \gamma}$,
and investigate a possible mild dependence of the probability that the photon interacts as a hadron ($P_{had}$)
increases with the energy. The results are compared to the data handled with the
PYTHIA and PHOJET Monte Carlo generators. The conclusions are drawn in Sec. V.

\section{A QCD-inspired eikonal model with a dynamical scale}

The high-energy hadron-hadron cross sections are calculated using a recently developed QCD-inspired eikonal
model, where the onset of the dominance of gluons in the interaction of high-energy hadrons is managed by the
dynamical gluon mass scale \cite{luna01}. The model satisfies analyticity and unitarity constraints, and the
hadron-hadron elastic scattering is due to the diffractive shadow of inelastic events. In the
eikonal formalism the total cross section is given by
\begin{eqnarray}
\sigma_{tot}(s) = 4\pi \int_{_{0}}^{^{\infty}} \!\! b\, db\, [1-e^{-\chi_{_{I}}(b,s)}\cos \chi_{_{R}}(b,s)],
\label{degt1}
\end{eqnarray}
where $s$ is the square of the total center-of-mass energy, $b$ is the impact parameter, and
$\chi(b,s)=\chi_{_{R}}(b,s)+i\chi_{_{I}}(b,s)$
is a complex eikonal function. Following the Ref. \cite{luna01}, we write the even eikonal as the sum of
gluon-gluon, quark-gluon, and quark-quark
contributions:
\begin{eqnarray}
\chi^{+}(b,s) &=& \chi_{qq} (b,s) +\chi_{qg} (b,s) + \chi_{gg} (b,s) \nonumber \\
&=& i[\sigma_{qq}(s) W(b;\mu_{qq}) + \sigma_{qg}(s) W(b;\mu_{qg})+ \sigma_{gg}(s) W(b;\mu_{gg})] ,
\label{final4}
\end{eqnarray}
where $\chi_{pp}^{\bar{p}p}(b,s) = \chi^{+} (b,s) \pm \chi^{-} (b,s)$. Here $W(b;\mu)$ is the overlap
function at impact parameter space and $\sigma_{ij}(s)$ are the elementary
subprocess cross sections of colliding quarks and gluons ($i,j=q,g$). The overlap function, normalized
so that $\int d^{2}\vec{b}\, W(b;\mu)=1$, is associated with the Fourier transform of a dipole form factor,
\begin{eqnarray}
W(b;\mu) = \frac{\mu^2}{96\pi}\, (\mu b)^3 \, K_{3}(\mu b),
\end{eqnarray}
where $K_{3}(x)$ is the modified Bessel function of second kind. The odd eikonal $\chi^{-}(b,s)$, that
accounts for the difference between $pp$ and $\bar{p}p$ channels, is parametrized as
\begin{eqnarray}
\chi^{-} (b,s) = C^{-}\, \Sigma \, \frac{m_{g}}{\sqrt{s}} \, e^{i\pi /4}\, 
W(b;\mu^{-}),
\label{oddeik}
\end{eqnarray}
where $m_{g}$ is the dynamical gluon mass and the parameters $C^{-}$ and $\mu^{-}$ are constants to be
fitted. The factor $\Sigma$ is defined as
\begin{eqnarray}
\Sigma = \frac{9\pi \bar{\alpha}_{s}^{2}(0)}{m_{g}^{2}},
\end{eqnarray}
with the dynamical coupling constant
$\bar{\alpha}_{s}$ set at its frozen infrared value. The origin of the dynamical gluon mass and the
frozen coupling constant can be traced back to the early work of Cornwall \cite{cornwall}, and the formal
expressions of these quantities can be seen in Ref. \cite{luna01}. 

Since that the main contribution to the asymptotic behavior of hadron-hadron total cross sections
comes from gluon-gluon semihard collisions, the eikonal functions $\chi_{qq} (b,s)$ and
$\chi_{qg} (b,s)$, needed to describe the lower-energy forward data, are simply parametrized with terms
dictated by the Regge phenomenology:
\begin{eqnarray}
\chi_{qq}(b,s) = i \, \Sigma \, A \,
\frac{m_{g}}{\sqrt{s}} \, W(b;\mu_{qq}),
\label{mdg1}
\end{eqnarray}
\begin{eqnarray}
\chi_{qg}(b,s) = i \, \Sigma \left[ A^{\prime} + B^{\prime} \ln \left( \frac{s}{m_{g}^{2}} \right) \right] \,
W(b;\sqrt{\mu_{qq}\mu_{gg}}),
\label{mdg2}
\end{eqnarray}
where $A$, $A^{\prime}$, $B^{\prime}$, $\mu_{qq}$ and $\mu_{gg}$ are fitting parameters. 

In the model the gluon eikonal term $\chi_{gg}(b,s)$ (see the expression (\ref{final4})) is written as
$\chi_{gg}(b,s)\equiv \sigma_{gg}^{D\!PT}(s)W(b; \mu_{gg})$, where
\begin{eqnarray}
\sigma_{gg}^{D\!PT}(s) = C^{\prime} \int_{4m_{g}^{2}/s}^{1} d\tau \,F_{gg}(\tau)\,
\hat{\sigma}^{D\!PT}_{gg} (\hat{s}) .
\label{sloh1}
\end{eqnarray}

Here $F_{gg}(\tau)$ is the convoluted structure function for pair $gg$, $\hat{\sigma}^{D\!PT}_{gg}(\hat{s})$ is
the subprocess cross section and $C^{\prime}$ is a fitting parameter. In the
above expression it is introduced the energy threshold $\hat{s}\geq 4m_{g}^{2}$ for the final state gluons,
assuming that these are screened gluons \cite{cornwall2}. The structure function $F_{gg}(\tau)$ is given by
\begin{eqnarray}
F_{gg}(\tau)=[g\otimes g](\tau)=\int_{\tau}^{1} \frac{dx}{x}\, g(x)\,
g\left( \frac{\tau}{x}\right),
\end{eqnarray}
where $g(x)$ is the gluon distribution function, adopted as \cite{gregores}
\begin{eqnarray}
g(x) = N_{g} \, \frac{(1-x)^5}{x^{J}},
\label{distgf}
\end{eqnarray}
where $J=1+\epsilon$ and $N_{g}=\frac{1}{240}(6-\epsilon)(5-\epsilon)...(1-\epsilon)$.
The correct analyticity properties of the model amplitudes is ensured by substituting $s\to se^{-i\pi/2}$
throughout Eqs. (\ref{mdg1}), (\ref{mdg2}) and (\ref{sloh1}). Other details of the model can be seen
in Ref. \cite{luna01}, and the fitting parameters are shown in Table \ref{parameters}, in the case of a dynamical
gluon mass $m_g = 400$ MeV, which is the preferred value for $pp$ and $\bar{p}p$ scattering.

\section{Photon-proton and photon-photon reactions}

The total cross section $\sigma_{tot}$, the ratio of the real to the imaginary part of the forward
scattering amplitude $\rho$, and the nuclear slope $B$ for proton-proton and
antiproton-proton collisions have been found to be well described by the QCD-inspired eikonal model discussed
in the last section, which we shall refer simply as the DGM model. Hence we fix the input parameters
to the eikonal model calculations by using the
data on $pp$ and $\bar{p}p$ reactions in order to make predictions for $\sigma^{\gamma p}$ and
$\sigma^{\gamma \gamma}$ using the same values of the parameters. These values are shown in the
Table \ref{parameters}.

The fact that at high energies a photon can fluctuate itself into a hadronic state open many theoretical
possibilities. For example, in the model of Block {\it et al.} \cite{gregores} the data on $\gamma p$
total photoproduction and the total hadronic $\gamma \gamma$ cross sections can be derived from the $pp$ and
$\bar{p}p$ forward amplitudes using vector meson dominance (VMD) and the additive quark model. We verify
that this procedure, valid for QCD inspired models with an arbitrary infrared scale, can also be applied
to our eikonal model with a dynamical gluon mass being necessary only minimal changes displayed in the
sequence. Thus, considering that a strongly interacting photon behaves as a system of
two quarks (the additive quark model), the even amplitude for $\gamma p$ scattering can be obtained after
the substitutions $\sigma_{ij} \rightarrow \frac{2}{3}\, \sigma_{ij}$ and
$\mu_{ij} \rightarrow \sqrt{\frac{3}{2}}\mu_{ij}$ in the eikonal of Eq. (\ref{final4}), resulting in
the following expression:
\begin{eqnarray}
\chi^{+}_{\gamma p}(b,s) = i \frac{2}{3}\left[ \sigma_{qq}(s)\, W\! \left( b;\sqrt{\frac{3}{2}}\mu_{qq} \right)
+ \sigma_{qg}(s)\, W\! \left(b;\sqrt{\frac{3}{2}}\mu_{qg}\right)
+ \sigma_{gg}(s)\, W\! \left(b;\sqrt{\frac{3}{2}}\mu_{gg}\right) \right] .
\label{hudrt1}
\end{eqnarray}

Note that in the work of Block {\it et al.} \cite{gregores} the above procedure is used to obtain
directly the physical eikonal $\chi^{\gamma p}(b,s)$, what means that the odd part of the eikonal that
describes $pp$ and $\bar{p}p$ scattering is neglected, and $\chi^{\gamma p}(b,s)\equiv \chi^{+}_{\gamma p}(b,s)$.
This is a good approximation at high energies, where the total cross sections $pp$ and $\bar{p}p$
are asymptotically expected to be the same. However, below $\sqrt{s} \sim 100$ GeV the $pp$ and $\bar{p}p$ total
cross sections are different, as happens in other channels like $\pi^-p$ and $\pi^+p$ as well as $K^-p$ and $K^+p$.
Therefore it is reasonable to expect the same behavior for the channels $\gamma p$ and $\gamma \bar{p}$. 
To account for these possible differences at low energy in the case of $\sigma^{\gamma p}$ and
$\sigma^{\gamma \bar{p}}$ we apply the prescription dictated by the additive quark model also in the odd eikonal 
of Eq. (\ref{oddeik}),
\begin{eqnarray}
\chi_{\gamma p}^{-}(b,s) =  \frac{2}{3}\, C^{-}\, \Sigma \, \frac{m_{g}}{\sqrt{s}} \, e^{i\pi /4}\,W\!
\left(b;\sqrt{\frac{3}{2}}\mu^{-}\right) ,
\label{eigp}
\end{eqnarray}
and the full physical eikonal, combining the even and odd parts, will be finally given by
$\chi^{\gamma \bar{p}}_{\gamma p}=\chi^{+}_{\gamma p}(b,s)\pm \chi^{-}_{\gamma p}(b,s)$.

Assuming vector meson dominance (VMD), the eikonalized $\gamma p$ total cross section can be written as
\begin{eqnarray}
\sigma^{\gamma p}_{tot}(s) = 4\pi \, P_{had}^{\gamma p} \int_{_{0}}^{^{\infty}} \!\! b\, db\,\rho\rho
[1-e^{-\chi^{\gamma p}_{_{I}}(b,s)}\cos \chi^{\gamma p}_{_{R}}(b,s)] ,
\label{degt2}
\end{eqnarray}
where $P_{had}^{\gamma p}$ is the probability that the photon interacts as a hadron. In the simplest VMD
formulation this probability is expected to be of $\mathcal{O}(\alpha_{em})$:
\begin{eqnarray}
P_{had}^{\gamma p} = P_{had} = \sum \limits_{V=\rho, \omega, \phi} 
\frac{4\pi\alpha_{em}}{f^2_V} \sim \frac{1}{249} ,
\label{phvmd}
\end{eqnarray}
where $\rho$, $\omega$ and $\phi$ are vector mesons. However, there are expected contributions to $P_{had}$ other
than $\rho$, $\omega$, $\phi$. For example, in the eikonalized minijet model analysis of $\sigma^{\gamma \gamma}$ of
the Ref. \cite{corsetti} a good fit to the photoproduction data is obtained with the value $P_{had}=1/204$, which
includes a non-VMD contribution of $\mathcal{O}(20\% )$. In this case there are strong contributions of heavier vector
mesons and continuum states. Moreover, the probability $P_{had}$ may also depend on the energy, which is a
possibility that we will explore in this paper.

To extend the model to the $\gamma \gamma$ channel we just perform the substitutions  
$\sigma_{ij}\to \frac{4}{9}\sigma_{ij}$ and
$\mu_{ij}\to \frac{3}{2} \mu_{ij}$ in the even part of the eikonal (Eq. (\ref{final4})):
\begin{eqnarray}
\chi^{\gamma \gamma}(b,s) = i \frac{4}{9} \left[ \sigma_{qq}(s)\, W\! \left(b;\frac{3}{2}\mu_{qq}\right)
+ \sigma_{qg}(s)\, W\! \left(b;\frac{3}{2}\mu_{qg}\right)
+ \sigma_{gg}(s)\, W\! \left(b;\frac{3}{2}\mu_{gg}\right) \right] ;
\label{hudrt2}
\end{eqnarray}
hence the additive quark model and the VMD are again assumed to be valid, and the calculation leads to the 
following eikonalized total $\gamma \gamma$ hadronic cross section 
\begin{eqnarray}
\sigma^{\gamma \gamma}_{tot}(s) = 4\pi \, P_{had}^{\gamma \gamma}
\int_{_{0}}^{^{\infty}} \!\! b\, db\, [1-e^{-\chi^{\gamma
\gamma}_{_{I}}(b,s)}\cos \chi^{\gamma \gamma}_{_{R}}(b,s)] 
\label{hudrt3}
\end{eqnarray}
where
\begin{eqnarray}
P_{had}^{\gamma \gamma} = P_{had}^{2} .
\label{phgg}
\end{eqnarray}

With the eikonal parameters of the DGM model fixed by the $pp$ and $\bar{p}p$ data (Table \ref{parameters}), 
we have performed all calculations of photoproduction and photon-photon scattering. 
Apart from the equations that we have displayed above, we will introduce two possible
modifications when we present our results in the next section. One is related
to a feasible energy dependence of $P_{had}$, and the other is concerned with the requirement
of a normalization factor in the $\sigma^{\gamma \gamma}$ calculation.

\section{results}

In our analysis the center-of-mass energy thresholds for the $\gamma p$ and $\gamma \gamma$ channels
are the same that one used in the study of the $pp$ and $\bar{p}p$ channels:
$\sqrt{s}^{\, min}_{\gamma p}= W_{\gamma \gamma}^{min} = 10$ GeV.
We adopt a $\chi^{2}$ fitting procedure, assuming an interval
$\chi^{2}-\chi^{2}_{min}$ corresponding, in the case of normal errors, to the projection of the
$\chi^{2}$ hypersurface containing 90\% of probability. In the case of 1, 2 and 3 free parameters,
this corresponds to the interval $\chi^{2}-\chi^{2}_{min}=2.70$, $4.61$ and $6.25$, respectively.
We use the data set for $\sigma^{\gamma p}$ compiled and analyzed by the Particle Data Group (PDG)
\cite{eidelman}, with the statistic and systematic errors added in quadrature. Data on $\gamma \gamma$
total hadronic cross sections are also available on the PDG compilation, but the $\gamma \gamma$
data quoted by the PDG are the average of the results obtained by unfolding the data with the PYTHIA
\cite{pythia} and the PHOJET \cite{phojet} Monte Carlo generators. These event generators, based on a model
by Schuler and Sj\"ostrand \cite{strojand} (PYTHIA) and on the Dual Parton
model \cite{dualpartmod} (PHOJET), are used to simulate photon-photon interactions and, in general, data unfolded
with PYTHIA are higher than if unfolded with PHOJET. As a consequence we obtain totally different scenarios
for the rise of the total hadronic $\gamma \gamma$ cross section at high energies. For this reason, we have
deconvoluted the data on $\sigma^{\gamma \gamma}$ of the L3 \cite{acciarri} and OPAL \cite{abbiendi}
collaborations, according to whether PYTHIA or PHOJET was used. Therefore, instead of using the PDG data for
$\sigma^{\gamma \gamma}$, we performed global fits considering separately the data obtained through
the PYTHIA and PHOJET codes, defining two data sets as
\begin{center}
SET I: $ \sigma^{\gamma p}$ and $\sigma^{\gamma \gamma}_{PYT}$ data 
($\sqrt{s}_{\gamma p}, W_{\gamma \gamma} \geq 10$ GeV), \\
SET II: $ \sigma^{\gamma p}$ and $\sigma^{\gamma \gamma}_{PHO}$ data 
($\sqrt{s}_{\gamma p}, W_{\gamma \gamma} \geq 10$ GeV),
\end{center}
where $\sigma^{\gamma \gamma}_{PYT}$ ($\sigma^{\gamma \gamma}_{PHO}$) correspond
to the data of $\gamma \gamma$ total hadronic cross section obtained via the
PYTHIA (PHOJET) generator. The unfolded results are depicted in the Figure 1. This
strategy of analysis is similar to the one adopted by Block and Kang, where a global test of
factorization for nucleon-nucleon, $\gamma p$ and $\gamma \gamma$ scattering was performed through
real analytic amplitudes \cite{block2}. Their results have evidenced the critical role played by the
Monte Carlo simulations used by L3 and OPAL group.

Our first analysis (henceforth referred to as FIT I) consisted in the determination of the $P_{had}^{-1}$
values from fits to the data sets defined above. In the fit to the SET I we obtained
$P_{had}^{-1} = 220.85 \pm 0.59$; the $\chi^2/DOF$ for this fit was 3.29. The fit to the SET II has resulted in
$P_{had}^{-1} = 223.70 \pm 0.50$, with $\chi^2/DOF=1.94$. These values are shown
in Table II. The $\sigma^{\gamma p}$ and $\sigma^{\gamma \gamma}$ cross sections
corresponding to these values of $P_{had}^{-1}$ are shown by the solid curves in Figure 2. 
The results of the fit to the SET I are displayed in
Figs. 2(a) and 2(b), whereas the results of the fit to the SET II are displayed in Figs. 2(c) and 2(d). In the FIT I
the $\chi^2/DOF$ values are relatively high. However, as we already commented, the extraction
of $\gamma \gamma$ data from $e^{+}e^{-}$ processes is strongly sensitive to the Monte Carlo simulation, and this
procedure could require a systematic overall-normalization factor in the experimental data. In fact, the
cross section $\sigma (\gamma \gamma\to hadrons)$ is determined from the
differential cross section of the reaction $e^{+}e^{-}\to e^{+}e^{-}\gamma^{*}\gamma^{*}\to e^{+}e^{-}hadrons$ by means
of both the luminosity function $L_{\gamma \gamma}$ for the photon flux and the hadronic cross section
$\sigma_{\gamma\gamma}(W_{\gamma\gamma},Q_{1}^{2},Q_{2}^{2})$ \cite{acciarri,abbiendi,budnev,acciarri2,wackerle}.
However, the uncertainty in the cross section extrapolation to real photons ($Q_{1}^{2}=Q_{2}^{2}=0$) is estimated
to be 5-7.5\%, and this uncertainty is not included in the systematic error of the measurement
\cite{wackerle}. In this case we
can introduce a normalization factor in the $\sigma^{\gamma \gamma}$ data as performed in Ref. \cite{block2} or, what
we have chosen to do, just multiply Eq. (\ref{hudrt3}) by a normalization factor $N$ to be fitted by the experiment:
\begin{eqnarray}
\sigma^{\gamma \gamma}_{tot}(s) = 4\pi \, N \, P_{had}^{\gamma \gamma}
\int_{_{0}}^{^{\infty}} \!\! b\, db\, [1-e^{-\chi^{\gamma
\gamma}_{_{I}}(b,s)}\cos \chi^{\gamma \gamma}_{_{R}}(b,s)] .
\label{hudrt4}
\end{eqnarray}

Within this approach (henceforth referred to as FIT II) we carried out new fits to the SETS I and II described above.
We obtained for the SET I (SET II) the values $P_{had}^{-1} = 222.30 \pm 1.18$ ($P_{had}^{-1} = 220.30 \pm 1.71$) and
$N=1.086\pm 0.028$ ($N=0.936\pm 0.025$), with $\chi^2/DOF=2.05$ ($\chi^2/DOF=1.02$). These values are shown
in Table II. The $\gamma p$ and $\gamma \gamma$ total cross sections
corresponding to these values of $P_{had}^{-1}$ and $N$ are shown by the dashed curves in Figure 2. 
The results of the fit to the SET I are displayed in
Figs. 2(a) and 2(b), whereas the results of the fit to the SET II are displayed in Figs. 2(c) and 2(d). We notice
that the approach adopted in the FIT II leads to an improvement of the fits, since the $\chi^2/DOF$ values obtained
through the FIT II are smaller than ones obtained through the FIT I. These results indicate the convenience of
introducing a normalization factor in order to account for possible effects not included in the systematic
error of $\sigma^{\gamma \gamma}$ data.

In the sequence we have also explored the possibility that $P_{had}$ varies with the energy. Such possibility
has been raised in the past \cite{durand5}, although there are arguments against 
it \cite{forshaw}. We have assumed a phenomenological expression for $P_{had}$, implying
that it increases logarithmically with the square of the center of mass energy:
$P_{had} = a + b \ln (s)$. Our intention was just to see if the quality of the fit
improves in such situation, referred to as FIT III. The results of the fit to the SET I were:
$a=(3.882\pm 0.122)\times 10^{-3}$, $b=(1.195\pm 0.229)\times 10^{-4}$ and $N=0.963\pm 0.037$, with $\chi^2/DOF=0.86$.
The fit results to the SET II were: $a=(4.175\pm 0.247)\times 10^{-3}$, $b=(6.272\pm 4.775)\times 10^{-5}$
and $N=0.875\pm 0.052$, with $\chi^2/DOF=0.76$. All these results are shown in Table III and the total cross section
curves are depicted in Figure 3, where Figs. 3(a) and 3(b) are related to the SET I whereas the Figs. 3(c) and 3(d)
are related to the SET II. Considering the smaller $\chi^2/DOF$ values obtained when
$P_{had}$ increases with the energy (FIT III), we believe that this possibility cannot be completely neglected.

In all the cases we see that the data of $\sigma^{\gamma \gamma}_{PYT}$ above
$\sqrt{s} \sim 100$ GeV can hardly be described by our model, but this is also a problem for other models in the
literature \cite{godbole7} based on the factorization assumption
\begin{eqnarray}
\frac{\sigma_{pp}}{\sigma_{\gamma p}} = \frac{\sigma_{\gamma p}}{\sigma_{\gamma \gamma}} .
\label{fact}
\end{eqnarray}

Assuming the correctness of these models we may interpret the steeper rise of the $\sigma^{\gamma \gamma}$ appointed
by the PYTHIA as a possible signal of violation of the factorization
relation. Otherwise this behavior indicates a possible failure of the PYTHIA generator at high-energies. The latter
possibility is qualitatively supported by the results depicted in the Figures 3(c) and 3(d), that show that the shape and
normalization of the PHOJET cross sections are in good agreement with the factorization relation (\ref{fact}).

\section{Conclusions}

Our first aim and main result was to show that an eikonal QCD 
inspired model based on a nonperturbative infrared 
dynamical gluon mass scale \cite{luna01}, with the help of vector meson dominance
and the additive quark model, can successfully describe the data
of the total photoproduction $\gamma p$ and total hadronic $\gamma \gamma$ cross sections. 
The calculation is similar to the one of Block {\it et al.} \cite{gregores}, with the
advantage that in the DGM model one has a physical meaning for the infrared scale, and
the coupling constant is not a free parameter, turning out to be related to the
dynamical gluon mass scale.
 
We promoted some minor changes in the calculation, assuming that $P_{had}$, the probability 
that a photon interacts as a hadron, has a logarithmic increase with $s$. This
leads to a improvement in the fit, i. e. it lowers the $\chi^2/DOF$ values. There are
pros \cite{durand5} and cons \cite{forshaw} about this possibility.
We just point out that the logarithmic increase of $P_{had}$ with $s$ is quite favored by the data.

We verified that the data of $\sigma^{\gamma \gamma}_{PYT}$ above
$\sqrt{s} \sim 100$ GeV can hardly be described by our
model, but this is also a problem for other models in the literature. Assuming
the correctness of these models we could say that the PHOJET generator is
more appropriate to obtain the $\sigma^{\gamma \gamma}$ data above 
$\sqrt{s} \sim 100$ GeV. Connected with this comment we can recall that
recently Block and Kaidalov demonstrated that the factorization 
relation (\ref{fact}) does not depend on the assumption of an 
additive quark model, but more on the opacity of the
eikonal being independent of the nature of the reaction \cite{kaidalov}. 
Hence, according to this result, we argue that it will be 
difficult for QCD-inspired models incorporating the total cross
section factorization hypothesis to accommodate the 
$\sigma (\gamma \gamma\to hadrons)$ data
with $\sqrt{s}_{\gamma \gamma}>100$ GeV unfolded with the PYTHIA
Monte Carlo generator.

\begin{acknowledgments}
We thank G. Azuelos for the reading of the manuscript.
%We thank G. Azuelos for criticism and useful discussions.
This research was supported by the Conselho
Nacional de Desenvolvimento Cient\'{\i}fico e Tecnol\'ogico-CNPq under contracts 151360/2004-9 and 301002/2004-5.
\end{acknowledgments}

\newpage

\begin{table*}
\caption{Values of the parameters of the DGM model resulting from the global fit to the forward $pp$ and
$\bar{p}p$ data \cite{luna01}. The dynamical gluon mass scale was set to $m_{g}=400$ MeV.}
\begin{ruledtabular}
\begin{tabular}{cc}
$C^{\prime}$ & (12.097$\pm$0.962)$\times 10^{-3}$ \\
$\mu_{gg}$ [GeV]& 0.7242$\pm$0.0172 \\
$A$ & 6.72$\pm$0.92 \\
$\mu_{qq}$ [GeV] & 1.0745$\pm$0.0405 \\
$A^{\prime}$ & (4.491$\pm$0.179)$\times 10^{-3}$ \\
$B^{\prime}$ & 1.08$\pm$0.14 \\
$C^{-}$ & 3.17$\pm$0.35 \\
$\mu^{-}$ [GeV]& 0.6092$\pm$0.0884 \\
\label{parameters}
\end{tabular}
\end{ruledtabular}
\end{table*}

\begin{figure}
\begin{center}
\includegraphics[height=.4\textheight]{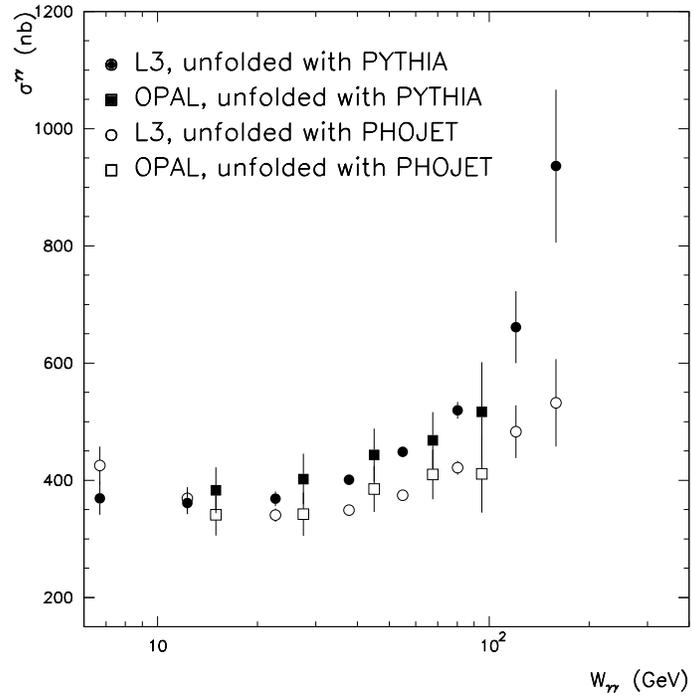}
\caption{$\sigma^{\gamma \gamma}$ data from the L3 and OPAL
collaborations unfolded with the PYTHIA and PHOJET generators.}
\end{center}
\label{oat3d1}
\end{figure}

\begin{table*}
\caption{Fits adopting $P_{had}$ constant.}
\begin{ruledtabular}
\begin{tabular}{ccccc}
 &  \multicolumn{2}{c}{FIT I} & \multicolumn{2}{c}{FIT II} \\
 &  SET I & SET II & SET I & SET II \\
\hline
$P_{had}^{-1}$ & 220.85$\pm$0.59 & 223.70$\pm$0.50 & 222.30$\pm$1.18 & 222.30$\pm$1.71 \\
$N$ & - & - & 1.086$\pm$0.028 & 0.936$\pm$0.025 \\
$\chi^2 /DOF$ & 3.29 & 1.94 & 2.05 & 1.02  \\
\hline
FIGURES  & 2(a), 2(b) & 2(c), 2(d) & 2(a), 2(b) & 2(c), 2(d) \\
\end{tabular}
\end{ruledtabular}
\end{table*}

\begin{figure}
\begin{center}
%\vspace{1.0cm}
\vglue 0.0cm
\hglue -8.0cm
\includegraphics[height=.35\textheight]{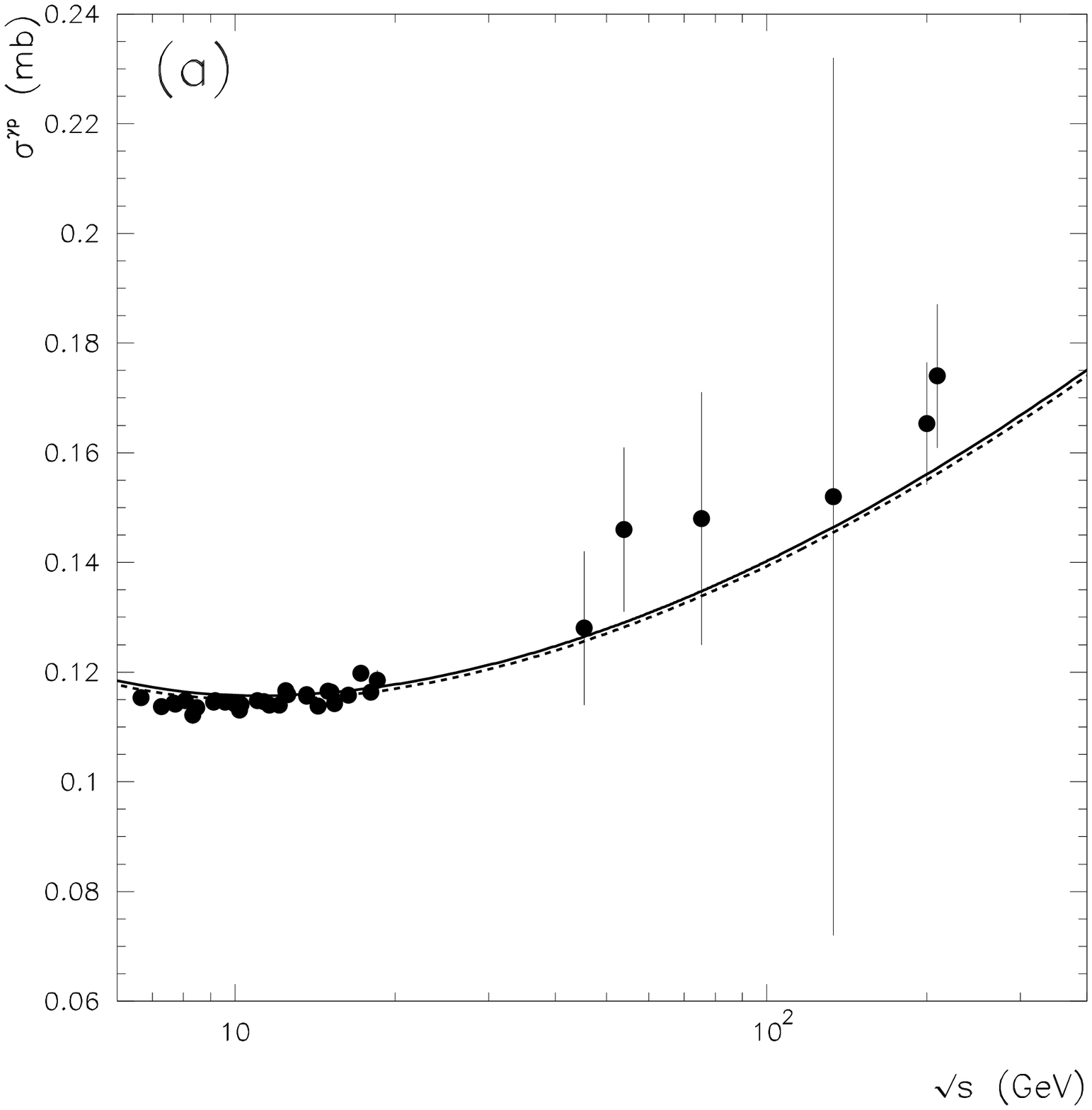}
\vglue -8.22cm
\hglue 8.3cm
\includegraphics[height=.35\textheight]{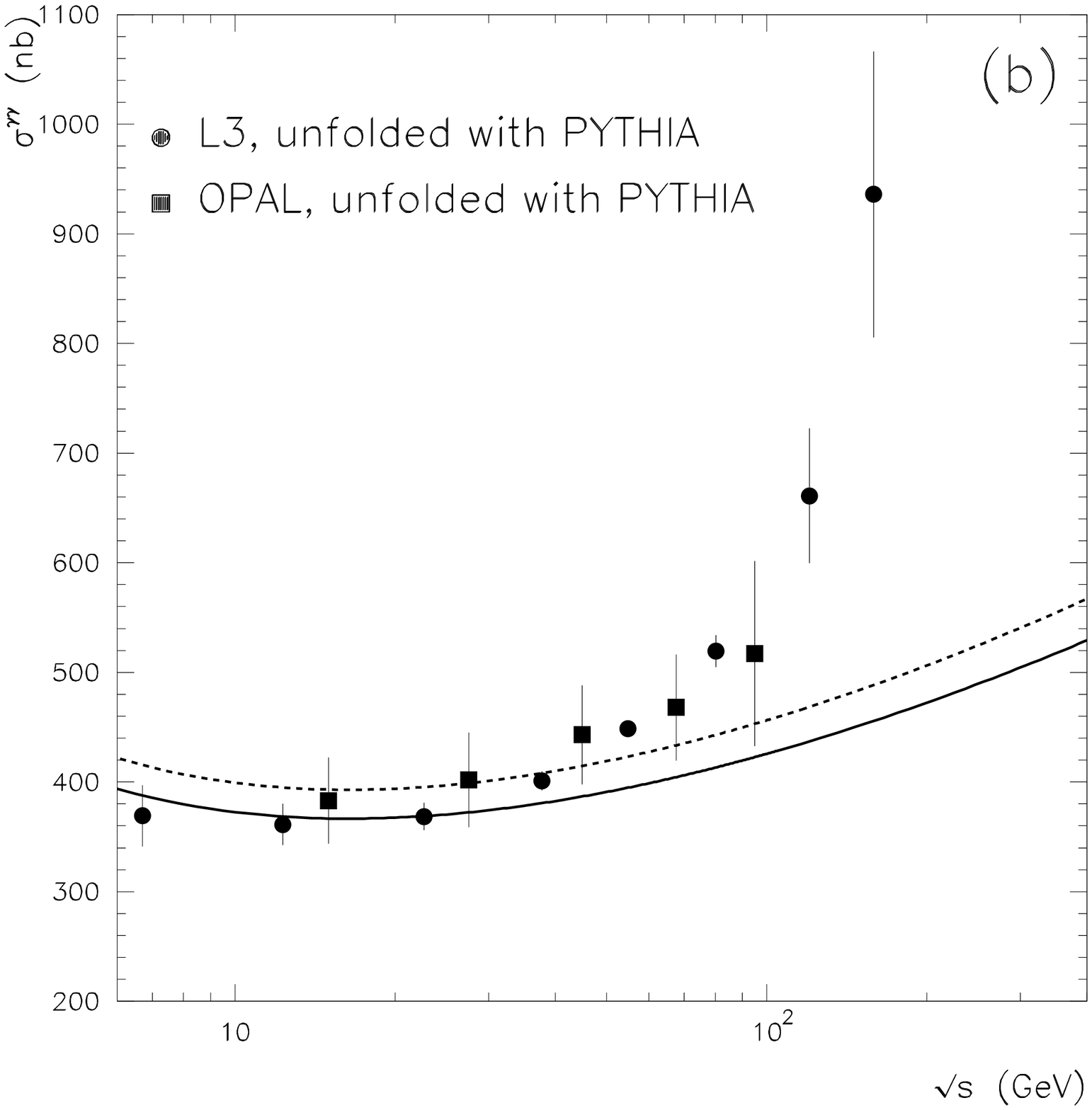}
\vglue 0.8cm
\hglue 8.3cm
\includegraphics[height=.35\textheight]{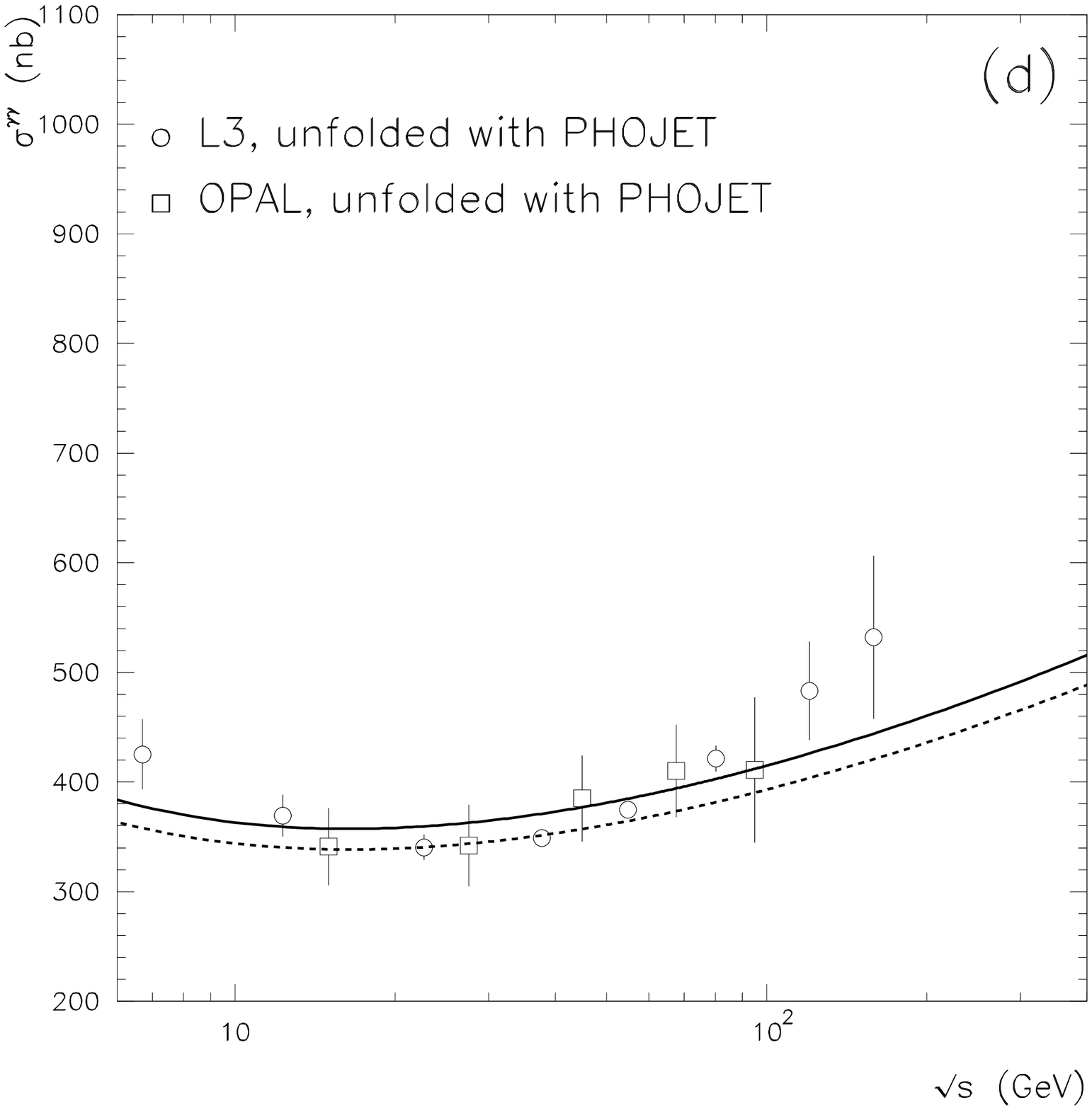}
\vglue -8.22cm
\hglue -8.0cm
\includegraphics[height=.35\textheight]{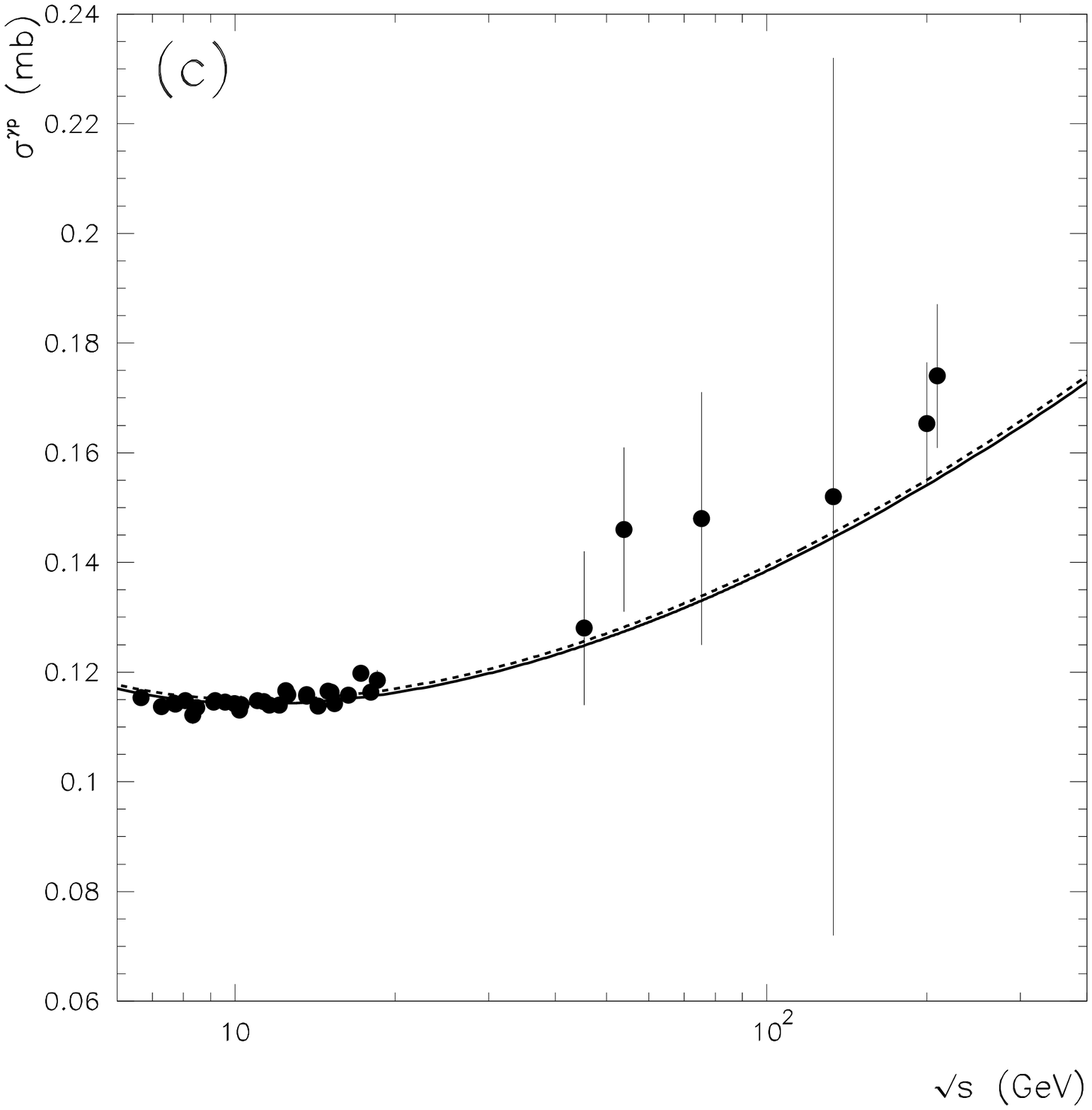}
\caption{$\sigma^{\gamma p}$ and $\sigma^{\gamma \gamma}$ cross sections
corresponding to the constant values of $P_{had}^{-1}$ shown in Table II. The
solid (dashed) curves correspond to the FIT I (II), whereas 
the results of the fit to the SET I are displayed in
Figs. 2(a) and 2(b), and the results of the fit to the SET II 
are displayed in Figs. 2(c) and 2(d).}
\end{center}
\label{frtd2}
\end{figure}

\begin{figure}
\begin{center}
%\vspace{1.0cm}
\vglue 0.0cm
\hglue -8.0cm
\includegraphics[height=.35\textheight]{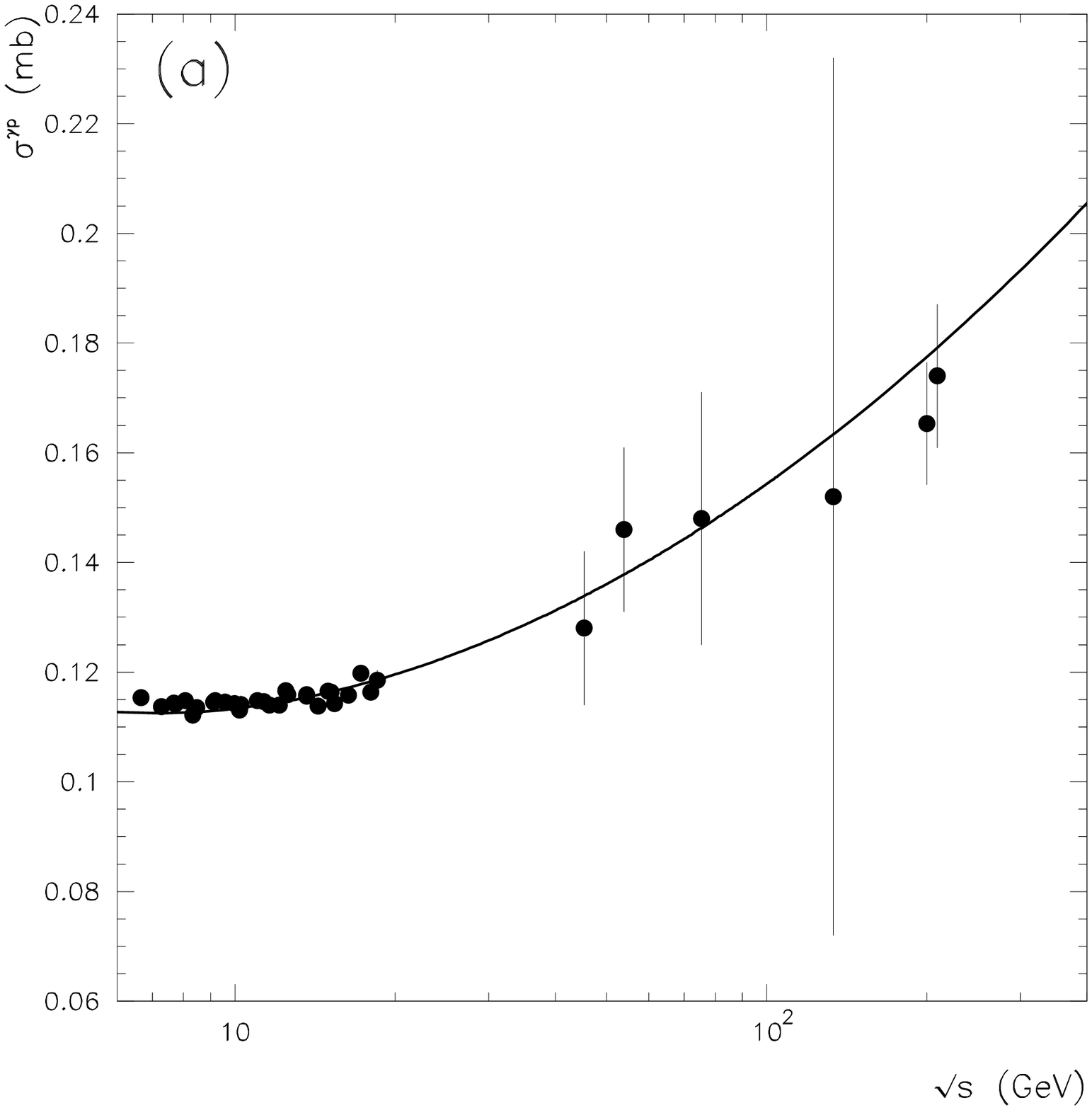}
\vglue -8.22cm
\hglue 8.3cm
\includegraphics[height=.35\textheight]{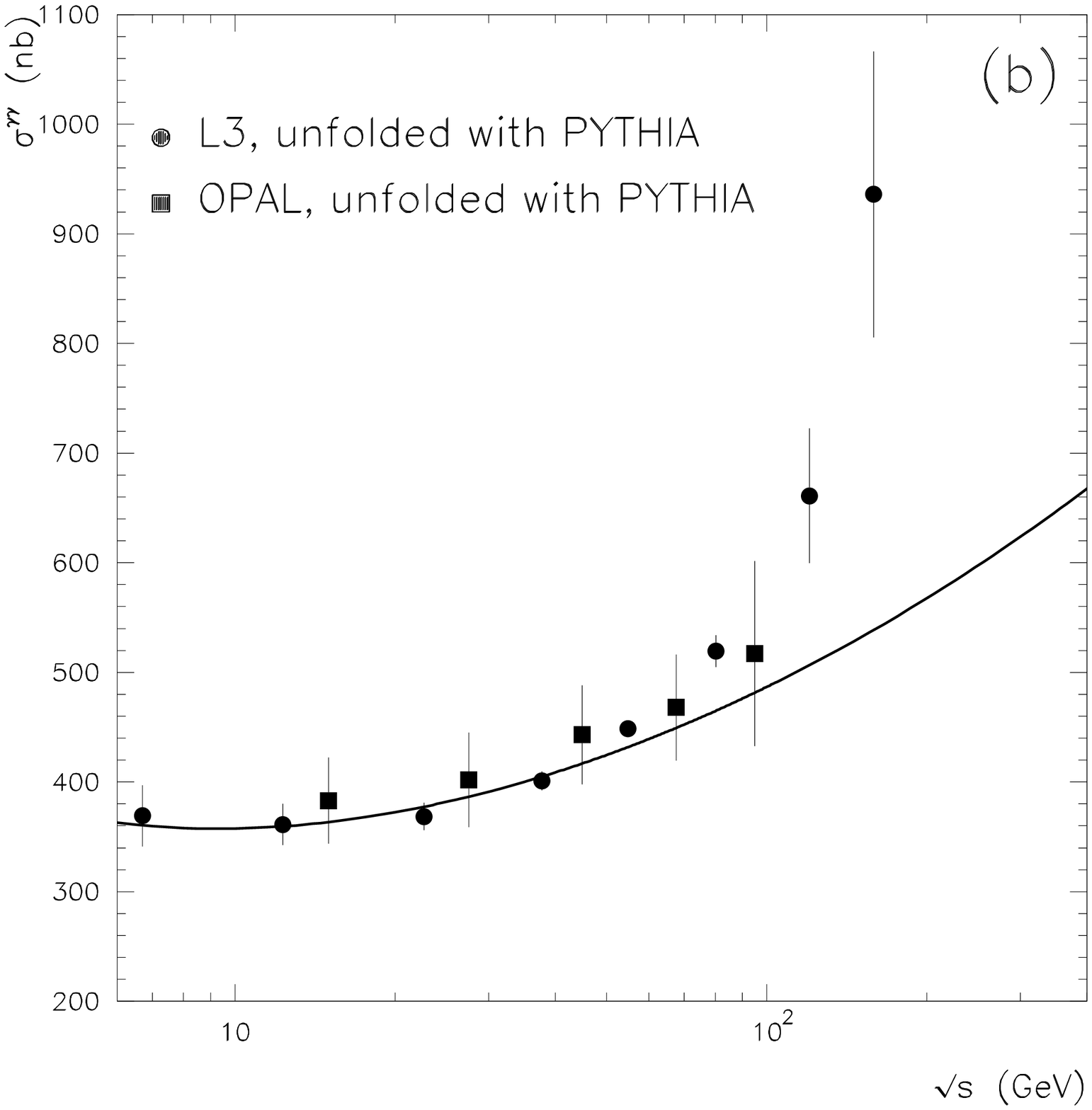}
\vglue 0.8cm
\hglue 8.3cm
\includegraphics[height=.35\textheight]{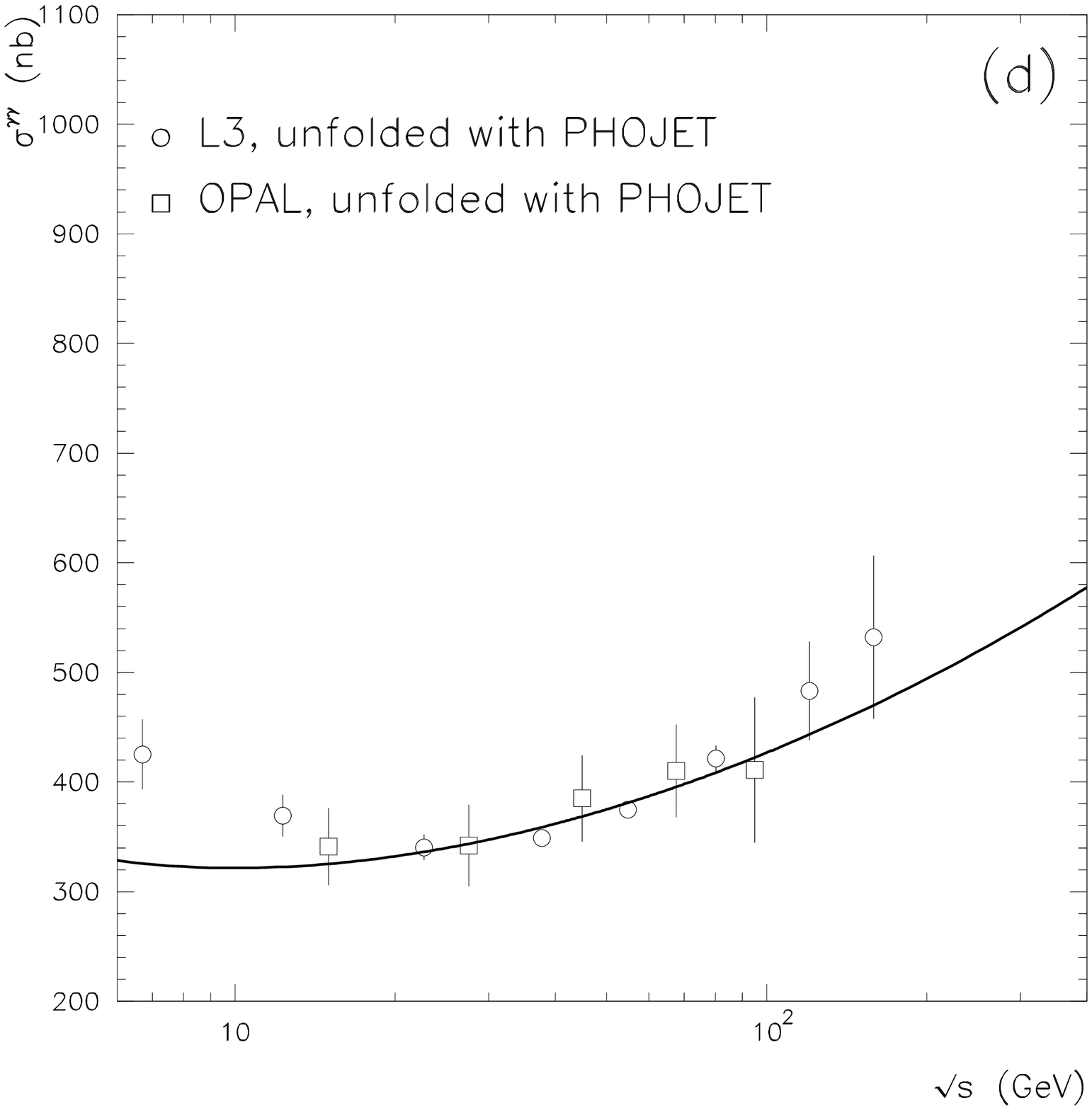}
\vglue -8.22cm
\hglue -8.0cm
\includegraphics[height=.35\textheight]{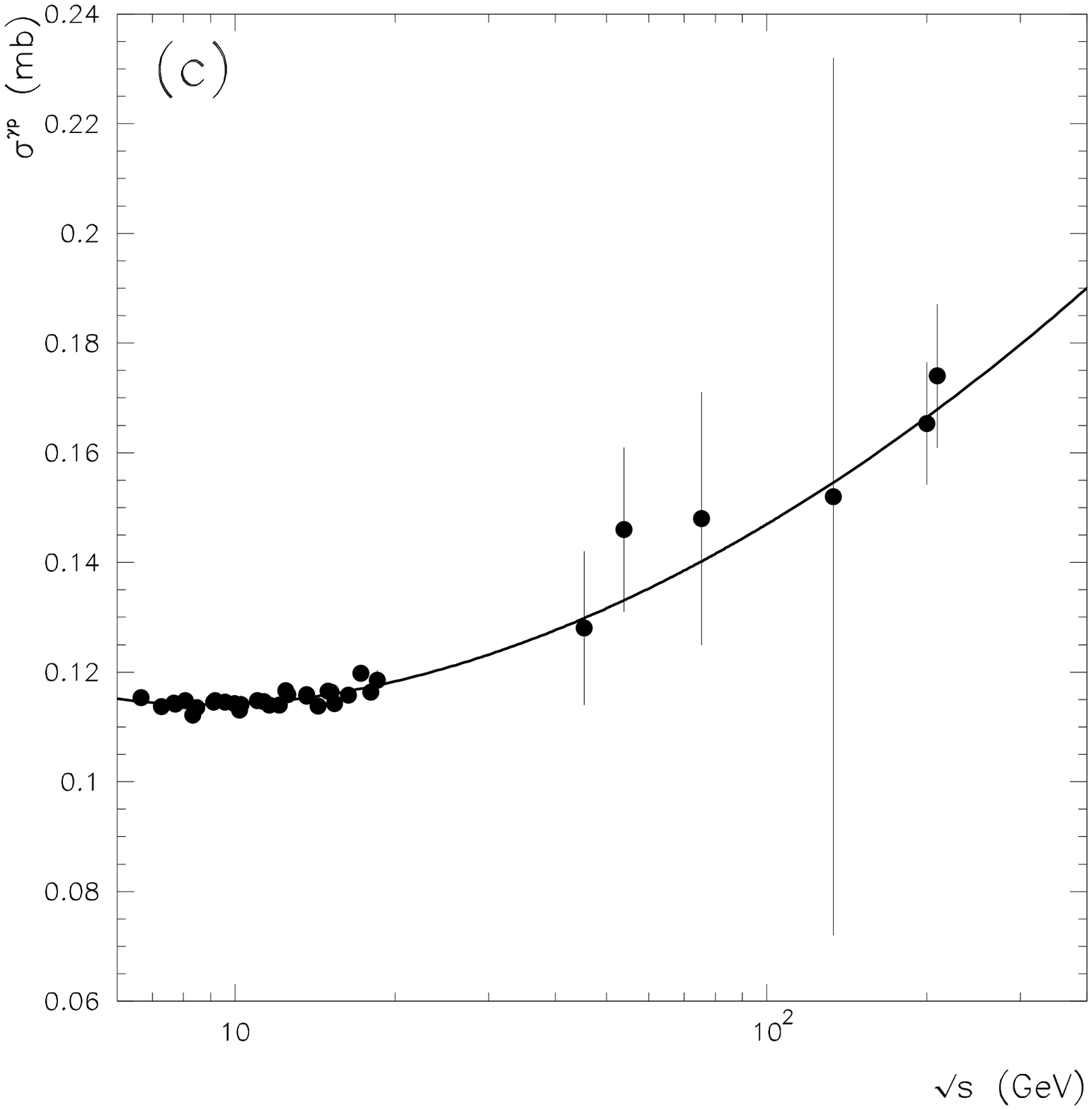}
\caption{$\sigma^{\gamma p}$ and $\sigma^{\gamma \gamma}$ cross sections
corresponding to the the case where $P_{had}^{-1}$ varies with the energy as
described in Table III. The curves of Figs. 3(a) and 3(b) are related to the fit
labeled as SET I in Table III, whereas the Figs. 3(c) and 3(d)
are related to the values of the SET II displayed in the same table. }
\end{center}
\label{frtd3}
\end{figure}

\begin{table*}
\caption{Fits adopting $P_{had}=a+b\ln (s)$.}
\begin{ruledtabular}
\begin{tabular}{ccc}
% & \multicolumn{2}{c}{Fit IV} \\
 & SET I & SET II \\
\hline
$a$ & (3.882$\pm$0.122)$\times 10^{-3}$ & (4.175$\pm$0.247)$\times 10^{-3}$ \\
$b$ & (1.195$\pm$0.229)$\times 10^{-4}$ & (6.272$\pm$4.775)$\times 10^{-5}$ \\
$N$ & 0.963$\pm$0.037 & 0.875$\pm$0.052 \\
$\chi^2 /DOF$ & 0.86 & 0.76 \\
\hline
FIGURES  & 3(a), 3(b) & 3(c), 3(d) \\ 
\end{tabular}
\end{ruledtabular}
\end{table*}

\end{document}